\documentclass[11pt]{article}
\usepackage{hyperref}
\pdfoutput=1
\begin{document}
\title{Dynamics of the Faraday Instability in a Small Cylinder}
\author{William Batson$^{1,2}$\thanks{wbatson@gmail.com} , Farzam Zoueshtiagh$^2$, Ranga Narayanan$^1$ \\
\\\vspace{6pt} $^1$University of Florida Department of Chemical Engineering \\
$^2$Universit\'e Lille 1 IEMN CNRS 8520}
\maketitle
\begin{abstract}
Vertical oscillation of a fluid interface above a critical amplitude excites the Faraday instability, typically manifesting itself as a standing wave pattern.  Fundamentally, the phenomenon is an example of parametric resonance.  At high frequencies, the wavelength is small and the pattern selection process is highly nonlinear.  We excite the instability with low frequencies, where the wavelength is large, and the form is highly influenced by the container geometry.  In this regime, the cell modes are easily excited on an individual basis and the observed waves resemble the forms predicted from linear theory.  In our video we highlight basic spatial and temporal dynamics of this regime.  This fluid dynamics video is submitted to the APS DFD Gallery of Fluid Motion 2013, part of the 66$^{th}$ Annual Meeting of the American Physical SocietyÕs Division of Fluid Dynamics (24-26 November, Pittsburgh, PA, USA).
\end{abstract}
\section{Video description}
After introduction, we first we present the growth and saturation of a (0,1) mode, and highlight its subharmonic response with respect to the imposed oscillation.  Next we show a (2,1) mode as an example of a different mode, and that saturation is not guaranteed as here the interfacial wave grows and breaks catastrophically.  At low frequencies, subharmonic response is not the only possibility, and we show examples of harmonic and superharmonic excitation for the (1,2) mode.  Finally we present an experiment where the parametric excitation is composed of two components, and three modes are excited and interact simultaneously.  Our experimental cylinder is 6.4 cm tall and 5.1 cm in diameter, we fill it completely with the fluids FC70 (1888 kg/m$^3$, ~12 cSt) and silicone oil (846 kg/m$^3$, ~1.5 cSt), and create an immiscible interface (tension ~7 dyne/cm) at a height of 3.1 cm.  The cell is mounted to an electromechanical shaker and oscillated at programmed frequencies and amplitudes--up to 15 Hz and accelerations of 3$g$, and the interface is recorded with a high speed camera.\\
\\\vspace{6pt}
For further information please see:
\begin{enumerate}
\item W. Batson, F. Zoueshtiagh, and R. Narayanan. ``The Faraday threshold in small cylinders and the sidewall non-ideality." \textit{Journal of Fluid Mechanics} 729 (2013): 496-523.
\end{enumerate}
\end{document}